# HANDOVER NECESSITY ESTIMATION FOR 4G HETEROGENEOUS NETWORKS

Issaka Hassane Abdoulaziz[1] , Li Renfa[2] and Zeng Fanzi[3]

[1,2,3]College of Information Science and Engineering, Hunan University, Changsha City, China
`fabismer@hotmail.com, scc_lrf@hnu.cn, zengfanzi@126.com`

## ABSTRACT

*One of the most challenges of 4G network is to have a unified network of heterogeneous wireless networks. To achieve seamless mobility in such a diverse environment, vertical hand off is still a challenging problem. In many situations handover failures and unnecessary handoffs are triggered causing degradation of services, reduction in throughput and increase the blocking probability and packet loss.*
*In this paper a new vertical handoff decision algorithm handover necessity estimation (HNE), is proposed to minimize the number of handover failure and unnecessary handover in heterogeneous wireless networks. we have proposed a multi criteria vertical handoff decision algorithm based on two parts: traveling time estimation and time threshold calculation. Our proposed methods are compared against two other methods: (a) the fixed RSS threshold based method, in which handovers between the cellular network and the WLAN are initiated when the RSS from the WLAN reaches a fixed threshold, and (b) the hysteresis based method, in which a hysteresis is introduced to prevent the ping-pong effect. Simulation results show that, this method reduced the number of handover failures and unnecessary handovers up to 80% and 70%, respectively.*

## KEYWORDS

*Handover failure, Handover necessity estimation HNE, Mobile Network, Time threshold, unnecessary handover, WLAN IEEE 802.11, 4G network,*

## 1. INTRODUCTION

In the 4G wireless environment, a mobile user is able to continue using the mobile device while moving from one point of attachment to another. Such process is called a handover, by which a mobile terminal keeps its connection active when it migrates from the coverage of one network access point to another [1]. Depending on the access network that each point of attachment belongs to, the handover can be either horizontal or vertical [1], [2]. A horizontal handover or intra-system handover takes place between PoA supporting the same network technology, e.g., two geographically neighboring BSs of a 3G cellular network. On the other side, a vertical handover or intersystem handover occurs between PoA supporting different network technologies, [3] e.g., an IEEE 802.11 AP and a 3G BS. An example of horizontal and vertical handovers is illustrated in Figure 1, where a horizontal handover happens between two WLANs and a vertical handover takes place between an AP of a WLAN and a BS of a UMTS cellular.
Vertical handovers are implemented across heterogeneous cells of access systems, which differ in several aspects such as bandwidth, data rate, frequency of operation, etc. The different characteristics of the networks involved make the implementation of vertical handovers more challenging as compared to horizontal handovers [4].

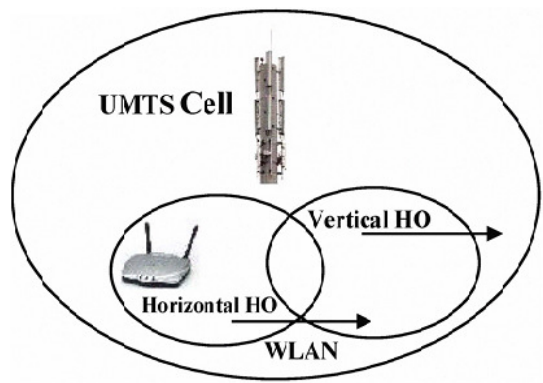

Figure 1: illustration of horizontal and vertical handoff.

Three stages are involved in a vertical handover process: network discovery, handover decision, and handover execution [5]. During the system discovery, mobile terminal equipped with multiple interfaces have to determine which networks can be used and what services are available in each network. During the handoff decision phase, the mobile device determines which network it should connect to. During the handoff execution phase, connections are needed to be re-routed from the existing network to the new network in a seamless manner .This requirement refers to the Always Best connected (ABC) concept, which includes the authentication ,authorization , as well as the transfer of user's context information [6].

This paper presents the vertical handoff management and focuses mainly on the handoff decision problem. Handover decision is the ability to decide when to perform the handover and to which access network to handover. A decision for vertical handover may depend on several issues relating to the network to which the mobile node is already connected and to the one that it is going to handover. For example, the decision to perform mobile-controlled handovers may be made by a vertical handover agent, sitting in the mobile device based on policies such as network bandwidth, load, coverage, cost, security, QoS, or even user preferences [7]. Several parameters have been proposed in the research literature for use in the vertical handover decision (VHD) algorithms. We briefly explain each of them below.

Handover delay: Handover delay is the duration between the initiation and completion of the handover process, and is related to the complexity of the VHD process. Reduction of the handover delay is especially important for delay sensitive voice or multimedia applications.

Number of handovers: Reducing the number of handovers is usually preferred as frequent handovers would cause wastage of network resources [8]. A handover is considered to be superfluous when a handover back to the original point of attachment is needed within certain time duration [9], and the number of such handovers should be minimized.

Handover failure probability: A handover failure occurs when the handover is initiated but the target network does not have sufficient resources to complete it, or when the mobile terminal moves out of the coverage of the target network before the process is finalized. In the former case, the handover failure probability is related to the channel availability of the target network [10], while in the latter case it is related to the mobility of the user [11].

Throughput: The throughput refers to the data rate delivered to the mobile terminals on the network. Handover to a network candidate with higher throughput is usually desirable.

Our proposed method (HNE) determines the necessity of making a handover to an available network. HNE takes various network parameters as its inputs and generates a binary value as its output. The inputs include: the AP power level, RSS samples, the radius of the network, the velocity of the MT, the handover latency, and the handover failure and unnecessary handover probability requirements. An output describes of '1' means a handover is necessary, and an output of '0' means the handover is not necessary. The block diagram of HNE is shown in Figure 2.

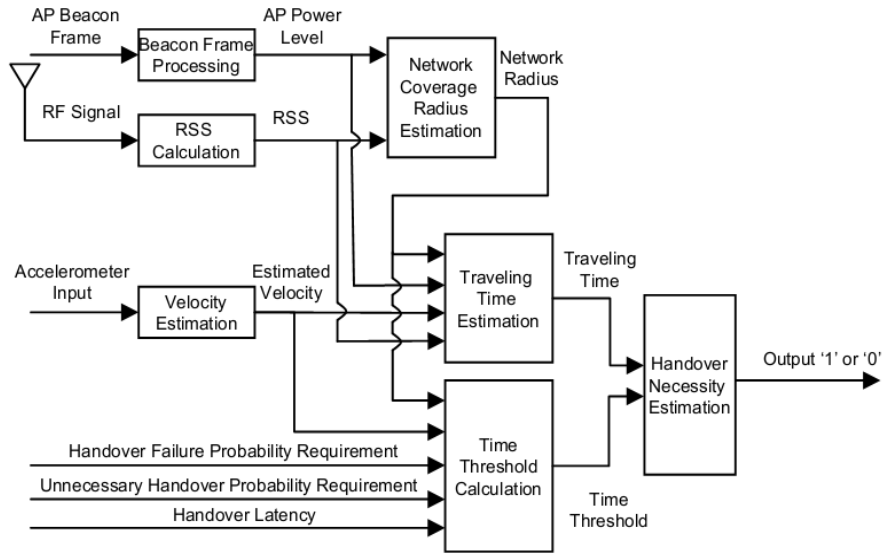

Figure 2: block diagram of handover necessity estimation (HNE)

The paper is structured as follows. Section 2 defines related work. The proposed algorithm for handover necessity estimation (HNE) is described in section 3. Section 4 present theoretical analyzes of HNE apply to previous works. Section 5 tackles the simulations to show the system performance with the proposed technique; to end with the paper conclusion in section 6.

## 2. RELATED WORK

A number of studies published earlier have surveyed VHD algorithms.

Zahran et al. [12] proposed an algorithm for handovers between 3G networks and WLANs by combining the RSS measurements either with an estimated life-time metric (expected duration after which the mobile terminal will not be able to maintain its connection with the WLAN) or the available bandwidth of the WLAN candidate. Benefits of Zahran et al.'s algorithm can be summarized as follows. First, by introducing the lifetime metric, the algorithm adapts to the application requirements and the user mobility, reducing the number of superfluous handovers significantly. Second, there is an improvement on the average throughput for the user because of the mobile terminal's ability to remain connected to the WLAN cell as long as possible. However, packet delays grow with an increase in the lifetime, due to the deterioration of the channel condition as the mobile terminal approaches the edge of the WLAN cell. This issue can be critical for delay sensitive applications and degrade their performance.

In [13] Li Jun Zhang et a. proposes a method to send probe requests to the APs one after the other and perform handoff immediately after any AP sends the response. This allows us to scan fewer channels. All these processes involve scanning of APs, it may be selective or all APs may be scanned. These methods are therefore time consuming as well as have a certain probability of handoff failure.

Yang et al. [14] presented a bandwidth based VHD method between WLANs and a Wideband Code Division Multiple Access (WCDMA) network using Signal to Interference and Noise Ratio (SINR). SINR based handovers can provide users with higher overall throughput than RSS based handovers since the available throughput is directly dependent on the SINR, and this algorithm results in a balanced load between the WLAN and the WCDMA networks. But such an algorithm may also introduce excessive handovers with the variation of the SINR causing the node to hand over back and forth between two networks, commonly referred to as ping-pong effect.

Mohanty [15] presented an algorithm for calculating a boundary area based on the speed of the MT and the WLAN cell size. In this algorithm, a handover from a WLAN to a 3G network is triggered when the MT enters the boundary area of the WLAN and handover procedures are completed before the MT leaves the WLAN. This algorithm operates efficiently for handovers from WLAN to 3G as it reduces the handover failure probability.

However, in the mobility architecture using this algorithm, and also in most of the other handover decision methods such as in [16], handovers from the cellular network to the WLAN are initiated once the MT enters the WLAN coverage area. This is not effective enough in situations where the MT travels through an area close to the coverage boundary of the WLAN at speeds above a certain threshold, since handovers to the WLAN become unnecessary. It is always better to avoid these handovers as much as possible since they lead to network resource wastage [8]. Furthermore, if the handover process has not been completed before the MT leaves the WLAN coverage area, connection breakdown inevitably occurs. In the method presented in [15] for handovers from the cellular network to the WLAN, the MT remains connected to both networks while staying in a boundary cell of the WLAN in order to avoid connection breakdown and also the ping-pong effect. However, this approach does not take into consideration the network resource wastage caused by unnecessary handovers. As yet, few studies on handover necessity estimation or on efficient methods for minimizing unnecessary handovers have been presented.

## 3. PROPOSED WORK

In this paper, a handover necessity estimation HNE method is introduced. This method is devised for minimizing handover failures and unnecessary handovers between cellular networks and WLANs, by estimating the necessity of a handover. Figure 2. The estimation involves two steps: time threshold calculation for minimizing handover failures and traveling time prediction for minimizing unnecessary handovers.

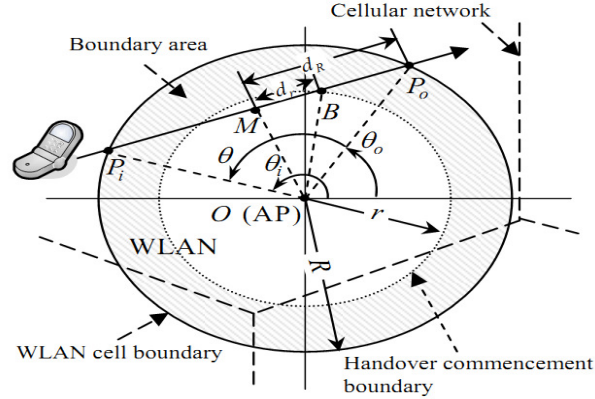

Figure 3: scheme diagram of WLAN to cellular network handover necessity estimation mechanism

Figure 3 shows the trajectory of a MT traveling over an area over which cellular network service is available and is also partially covered with a WLAN cell. The MT enters and exits the WLAN cell at points Pi and Po, respectively, following a straight line. M is the middle point of the section of the trajectory inside the WLAN cell. The donut shaped (dashed) area is called "boundary area". The radii of the outer and inner circles enclosing the boundary area are R and r, and dR and dr represent the half length of the trajectory segments inside the outer and inner circles respectively. B is the intersection point of the trajectory and the inner circle.

### 3.1. Time Threshold Calculation for Minimizing Handover Failures

The purpose of the time threshold calculation presented in this section is to keep the number of handover failures under a desirable threshold. That is, for example, if the system designer has a requirement of limiting the probability of handover failures under 1%, then the time threshold is adjusted to make the ratio of the number of failed handovers to the total number of handovers below 1%. The time threshold is calculated using mathematical modeling and probability calculation as explained below.

It is assumed that the entry and exit points Pi and Po can be any arbitrarily chosen points on the circle enclosing the WLAN coverage area, with equal probability (Figure 3). Then the angles $\theta_i$ and $\theta_o$ are both uniformly distributed in [0, 2π], and $\theta = |\theta_i - \theta_o|$.

The first step is to calculate the probability density function (PDF) [17] of $\theta$.
The PDFs of the locations of Pi and Po are given, respectively,

$$f(\Theta_i, \Theta_o) = \begin{cases} \frac{1}{4\pi^2}, & 0 \leq \Theta_i, \Theta_o \leq 2\pi, \\ 0, & otherwise, \end{cases} \quad (1)$$

$$fp_o(\Theta_o) = \begin{cases} \frac{1}{2\pi}, & 0 \leq \Theta_o \leq 2\pi, \\ 0, & otherwise, \end{cases} \quad (2)$$

Since the locations of Pi and Po are independent from each other, their joint PDF is given by

International Journal of Information Sciences and Techniques (IJIST) Vol.2, No.1, January 2012$$f(\Theta_i, \Theta_o) = \begin{cases} \frac{1}{4\pi^2}, & 0 \leq \Theta_i, \Theta_o \leq 2\pi, \\ 0, & otherwise, \end{cases} \quad (3)$$

The probability that, which is also the cumulative distribution function (CDF) of θ, can be derived using the following integral [18]

$$F(\Theta) = P(\theta \leq \Theta)$$
$$= \iint_\Omega f(\theta_i, \theta_o) d\theta_o d\theta_i, \quad (4)$$

Where $\Omega$ is the space of locations of entry and exit points Pi and Po such that $\theta \leq \Theta$ and $0 \leq \Theta \leq 2\pi$. $P(\theta \leq \Theta) = 0$ for $\Theta \leq 0$ and $P(\theta \leq \Theta) = 1$ for $\Theta > 2\pi$. From the observation of Figure 2 Equation (4) can be rewritten as

$$F(\Theta) = P(\theta \leq \Theta)$$
$$= 4\pi\Theta - 4\Theta^2, \quad 0 \leq \Theta \leq 2\pi. \quad (5)$$

The PDF of θ can be derived by taking the derivative of Equation (5) and is given by

$$f(\Theta) = \begin{cases} \frac{1}{\pi}(1 - \frac{\Theta}{2\pi}), & 0 \leq \Theta \leq 2\pi, \\ 0, & otherwise. \end{cases} \quad (6)$$

The next step is to use the PDF of θ and the expression of the traveling time $t_{WLAN}$ as a function of θ to obtain the PDF of $t_{WLAN}$.

From the geometric configuration in Figure 3 and by using the cosine formula, the following equation is obtained:

$$(vt_{WLAN})^2 = 2R^2(1 - COS\theta). \quad (7)$$

Thus,

$$t_{WLAN} = g(\theta)$$
$$= \sqrt{\frac{2R^2}{v^2}(1 - \cos\theta)}. \quad (8)$$

Using the theorem stated in [17], the PDF of $t_{WLAN}$ is expressed as

$$f(T) = \sum_1 n \frac{f(\theta_n)}{|g'(\theta_n)|}, \quad (9)$$

Where $\theta_1, ..., \theta_n$ are the root of function g(θ), and g'(.) is the derivative of g(.).
In equation (8), for g(θ) there are two roots, $\theta_1$ and $\theta_2$, which are expressed as

$$\theta_1 = arcos(1 - \frac{v^2 t_{WLAN}^2}{2R^2}), \quad (10)$$

$$\theta_2 = 2\pi - \arccos(1 - \frac{v^2 t_{WLAN}^2}{2R^2}), \quad (11)$$

From (8), g'(θ) is expressed as

$$g'(\theta) = \frac{R\sin\theta}{v\sqrt{2(1 - \cos\theta)}}. \quad (12)$$

$$|g'(\theta_1)| = \left| \frac{R\sin(arccos(1 - \frac{v^2 t_{WLAN}^2}{2R^2}))}{v\sqrt{2\left[1 - \cos(arccos(1 - \frac{v^2 t_{WLAN}^2}{2R^2}))\right]}} \right|$$

$$= R\sqrt{1 - \frac{v^2 t_{WLAN}^2}{4R^2}} \quad (13)$$

$$|g'(\theta_2)| = \left| \frac{R\sin(2\pi - \arccos(1 - \frac{v^2 t_{WLAN}^2}{2R^2}))}{v\sqrt{2\left[1 - \cos(2\pi - \arccos(1 - \frac{v^2 t_{WLAN}^2}{2R^2}))\right]}} \right|$$

$$= R\sqrt{1 - \frac{v^2 t^2}{4R^2}} \quad (14)$$

And

$$f(\theta_1) = \frac{1}{\pi}\left[1 - \frac{\arccos(1 - \frac{v^2 t_{WLAN}^2}{2R^2})}{2\pi}\right] \quad (15)$$

$$f(\theta_2) = \frac{1}{\pi}\left[1 - \frac{2\pi - \arccos(1 - \frac{v^2 t_{WLAN}^2}{2R^2})}{2\pi}\right] \quad (16)$$

Thus using equations (9), (13) and (15) the PDF of $t_{WLAN}$ is calculated by

$$f(T) = \begin{cases} \frac{f(\theta_1)}{|g'(\theta_1)|} + \frac{f(\theta_2)}{|g'(\theta_2)|}, & 0 \leq T \leq \frac{2R}{v} \\ 0, & otherwise. \end{cases}$$

$$= \begin{cases} \frac{2}{\pi\sqrt{4R^2 - v^2 T^2}}, & 0 \leq T \leq \frac{2R}{v}, \\ 0, & otherwise. \end{cases} \quad (17)$$

The third step is to use the PDF of $t_{WLAN}$ to obtain the CDF of $t_{WLAN}$, which is derived from the integral of Equation (17) as:

$$F(T) = \Pr[t \leq T]$$
$$= \int_0^T f(T)dT$$
$$= \begin{cases} 1, & \frac{2R}{v} \prec T, \\ \frac{2}{\pi}\arccos(\frac{vT}{2R}), & 0 \leq T \leq \frac{2R}{v}. \end{cases} \quad (18)$$

A time threshold parameter T1 is introduced to make handover decisions: whenever the estimated traveling time $t_{WLAN}$ is greater than T1, the MT will initiate the handover procedure. A handover failure occurs when the traveling time inside the WLAN cell is shorter than the handover latency from the cellular network to the WLAN, $T_i$. Thus, using Equation (18) the probability of a handover failure for the method using the threshold T1 is given by

$$P_f = \begin{cases} \frac{2}{\pi}\left[\arcsin(\frac{v\tau_i}{2R}) - \arcsin(\frac{vT1}{2R})\right], & 0 \leq T1 \leq \tau_i, \\ 0, & \tau_i \prec T1. \end{cases} \quad (19)$$

By using (19), an equation which can be used by the MT to calculate the value of T1 for a particular value of $P_f$ when $0 < P_f < 1$:

$$T1 = \frac{2R}{v}\sin(\arcsin(\frac{v\tau_i}{2R}) - \frac{\pi}{2}P_f) \quad (20)$$

To calculate T1, the speed of the MT $v$ and the handover latency $\tau_i$ need to be obtained. In this research, the knowledge of $v$ and $\tau_i$ is assumed, and they can be measured by using accelerometers [19] and the technique described in [20], respectively.

## 3.2. Traveling Time Prediction for Minimizing Unnecessary Handovers

To eliminate unnecessary handovers, Yan et al. [9] developed a VHD algorithm that takes into consideration the time the mobile terminal is expected to spend within a WLAN cell.
The method relies on the estimation of WLAN traveling time (i.e. time that the mobile terminal is expected to spend within the WLAN cell) and the calculation of a time threshold ($t_{WLAN}$). A handover to a WLAN is triggered if the WLAN coverage is available and the estimated traveling time inside the WLAN cell is larger than the time threshold. The estimated traveling time ($t_{WLAN}$) is

$$t_{WLAN} = \frac{R^2 - l_{OS}^2 + v^2(t_s - t_{p_i})^2}{v^2(t_s - t_{in})} \qquad (21)$$

Where R is the radius of the WLAN cell, $l_{OS}$ is the distance between the access point and where the mobile terminal takes an RSS sample, $v$ is the velocity of the mobile terminal, $t_s$ and $t_{p_i}$ are the times at which the RSS sample is taken and the mobile terminal enters the WLAN cell coverage, respectively. $l_{OS}$ is estimated by using the RSS information and log-distance path loss model.

The time threshold ($T_{WLAN}$) is calculated based on various network parameters as
$$T_{WLAN} = \frac{2R}{v}\sin(\sin^{-1}(\frac{v\tau}{2R}) - \frac{\pi}{2}P) \qquad (22)$$

where $\tau_i$ is the handover delay from the cellular network to the WLAN, and $P$ is the tolerable handover failure, unnecessary handover or connection breakdown probability. A handover to the cellular network is initiated if the WLAN RSS is continuously fading and the mobile terminal reaches a handover commencement boundary area which size is dynamic to the mobile terminal's speed.

Similar to the arguments used in section 3.1, another parameter T2 (T1< T2) is introduced to minimize the probability of unnecessary handovers. By using (18) the probability of an unnecessary handover is calculated as

$$P_u = \begin{cases} \frac{2}{\pi}\left[\arcsin(\frac{v(\tau_i+\tau_o)}{2R}) - \arcsin(\frac{vT2}{2R})\right], & 0 \leq T2 \leq (\tau_i+\tau_o), \\ 0, & (\tau_i+\tau_o) \prec T2. \end{cases} \qquad (23)$$

Thus
$$T2 = \frac{2R}{v}\sin(\arcsin(\frac{v(\tau_i+\tau_o)}{2R}) - \frac{\pi P_u}{2}) \qquad (24)$$

Equation (24) is derived from (23) for a particular value of $P_u$ when $0 < P_u < 1$.

Parameters T1 and T2 depend on values of constants $P_f$ and $P_u$ which are selected by system designers. They also depend on measurement of $v$, R, $\tau_i$ and $\tau_o$. The parameter T2 can be further adjusted dynamically to encourage or discourage handovers to WLAN by considering other performance criteria such as network load.

## 4. THEORETICAL ANALYSIS OF HNE FOR FIXED RSS THRESHOLD [16] AND HYSTERESIS BASED METHODS [21]

### 4.1. Handover Failure Probability

Using equation (18), the handover failure probability for the fixed RSS threshold based method is given by

$$P_{f_{fixed}} = \begin{cases} 1, & v\tau_i > 2R1_{fixed}, \\ \frac{2}{\pi}\sin^{-1}(\frac{v\tau_i}{2R1_{fixed}}), & 0 \leq v\tau_i \leq 2R1_{fixed} \end{cases} \quad (25)$$

where $R1_{fixed}$ is the distance between the MT location and the AP of the WLAN cell when a handover into the WLAN occurs in the fixed RSS threshold based method. It is calculated by

$$R1_{fixed} = 10^{\frac{E_t - RSS1_{fixed}}{10\beta}} \quad (26)$$

Using Equation (18), the handover failure probability for the hysteresis based method is given by

$$P_{f_{hyst}} = \begin{cases} 1, & v\tau_i > 2R1_{hyst}, \\ \frac{2}{\pi}\sin^{-1}(\frac{v\tau_i}{2R1_{hyst}}), & 0 \leq v\tau_i \leq 2R1_{hyst} \end{cases} \quad (27)$$

where $R1_{hyst}$ is the distance between the MT location and the AP of the WLAN cell when a handover into the WLAN occurs in the hysteresis based method. It is calculated by

$$R1_{hyst} = 10^{\frac{E_t - RSS1_{hyst}}{10\beta}} \quad (28)$$

### 4.2. Unnecessary Handover Probability

Using equation (18), the unnecessary handover probability for the fixed RSS threshold based method is given by

$$P_{u_{fixed}} = \begin{cases} 1, & v(\tau_i + \tau_o) > 2R1_{fixed}, \\ \frac{2}{\pi}\sin^{-1}(\frac{v(\tau_i+\tau_o)}{2R1_{fixed}}), & 0 \leq v(\tau_i + \tau_o) \leq 2R1_{fixed} \end{cases} \quad (29)$$

The unnecessary handover probability for the hysteresis based method is given by

$$P_{u_{hyst}} = \begin{cases} 1, & v(\tau_i + \tau_o) > 2R1_{hyst}, \\ \frac{2}{\pi}\sin^{-1}(\frac{v(\tau_i+\tau_o)}{2R1_{hyst}}), & 0 \leq v(\tau_i + \tau_o) \leq 2R1_{hyst} \end{cases} \quad (30)$$

## 5. SIMULATION

Our proposed methods are compared against two other methods: (a) the fixed RSS threshold based method [16], in which handovers between the cellular network and the WLAN are initiated when the RSS from the WLAN reaches a fixed threshold, and (b) the hysteresis based method [21], in which a hysteresis is introduced to prevent the ping-pong effect.

### 5.1. Evaluation Parameter

The parameters used in theoretical analysis and simulations of HNE are listed in Table 1.

**Table 1**: Parameters used in the HNE performance evaluation.

| Parameter | Symbol | Value |
|---|---|---|
| WLAN radius | R | 150 m |
| AP transmit power | $P_{TX}$ | 20 dBm |
| Distance between the AP and the reference point | $d_{ref}$ | 1m |
| Path loss at the reference point | $PL_{ref}$ | 40dB |
| Path loss exponent | β | 3.5 |
| Standard deviation of shadow fading | σ | 4.3dB |
| Handover delay from cellular network to WLAN | $\Gamma_i$ | 2S |
| Handover delay from WLAN to cellular network | $\Gamma_o$ | 2S |
| Tolerable handover failure probability | $P_f$ | 0.02 |
| Tolerable unnecessary handover probability | $P_u$ | 0.04 |

MATLAB was used for the experiments, which generated 10000 random MT trajectories across the WLAN cell coverage area for speeds from 3.6 km/h to 100 km/h in 2 km/h increments. For each trajectory, a random WLAN cell entry point was chosen, and a uniformly distributed random angle between 0 and 2π was generated representing the movement direction of the MT.

### 5.2. Simulation Analysis

Two critical parametric quantities are examined: number of handover failure and unnecessary handover.

The number of handover failures and unnecessary handovers of the RSS threshold based ($R1_{fixed}$ = 150 m), hysteresis based ($R1_{hyst}$ = 120 m) and HNE methods are shown in Figures 4 and 5.

From the figures it can be seen that, with HNE, handover failures and unnecessary handovers are kept under the numbers of 200 and 500, respectively. The total number of handovers declines with the increasing velocity of the MT. In the RSS threshold and hysteresis based methods, the numbers of handover failures and unnecessary handovers increase sharply as the velocity increases. HNE is able to reduce the number of handover failures and unnecessary handovers up to 80%, when the velocity of the MT is up to 100 km/h. HNE yields much better performance than the other two methods. Otherwise, for velocities less than 20 km/h, the other two methods yield marginally better results.

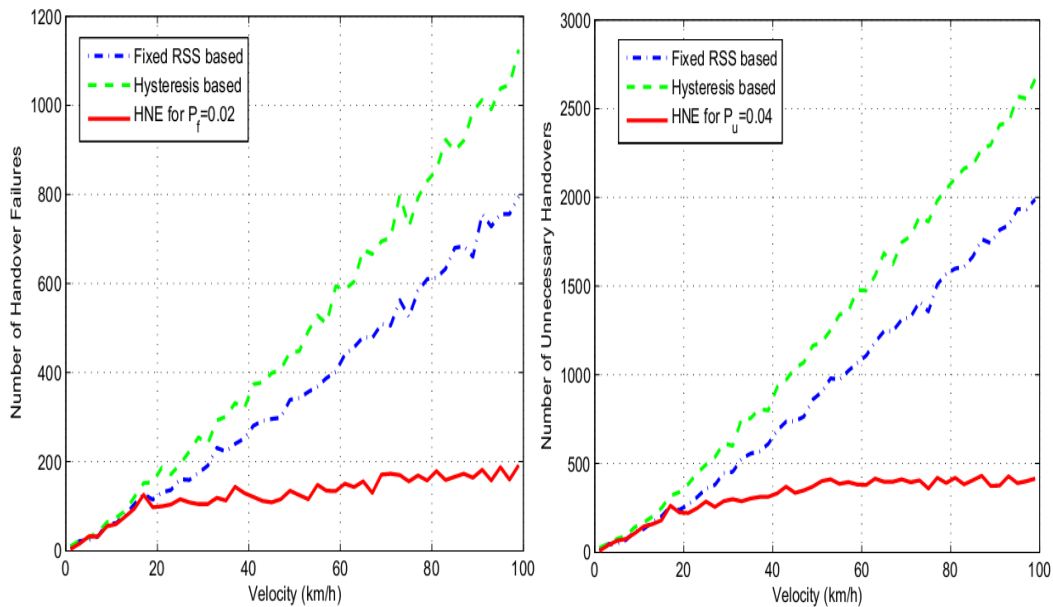

Fig 4: Number of handover failures of [16], [21] and HNE methods.  Fig 5: unnecessary handover of [16], [21] and HNE

## 6. CONCLUSION

In this paper, a method to estimate the handover necessity into a WLAN cell is presented. This method is based on two parts: traveling time estimation and time threshold calculation. The traveling time estimation relies on the RSS measurements and the speed of the MT. The time thresholds are calculated based on various network parameters such as tolerable handover failure probability or un-necessary handover probability, the radius of the WLAN cell and the handover latency. This method is able to reduce the number of handover failures and unnecessary handovers up to 80% and 70%, comparing with the conventional RSS threshold based [16] and hysteresis based [21].

### ACKNOWLEDGEMENTS

This work was supported in part by the national natural science foundation of china with grant number 60903019 and the science and technology planning  project of Hunan province; China. No. 2010GK3051

## Authors

**Issaka Hassane Abdoulaziz** received the B.S. degree in Networks and Telecommunications from EST-LOKO institute, Abidjan, Cote d'Ivoire, in June 2004 and the M.S. in computer science and Technology from Hunan University, Hunan, China, in July 2009. He is currently working toward his Ph.D. degree in the college of information science and engineering at Hunan University, China.
From 2004 to 2005 he worked in the Ministry of National Education at the Human Resource Department, Niger. He worked for the High Council of Communication, Niger, from 2005 to 2006. His research interests include wireless network, mobile communications, cognitive radio, and orthogonal frequency-division multiplexing.

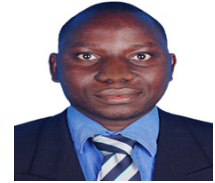

**Li Renfa** born in 1956. Professor and Ph.D supervisor in Hunan University. His main research interests are embedded system and network.

**Zeng Fanzi** born in 1971. Associate professor, Hunan University. His main research interests include UWB and cognitive radio
hort Biography